\documentclass[11pt,a4paper,reqno]{article}
\usepackage{mysty}
\usepackage{fullpage}

\usepackage[thinlines]{easytable}

\SetLabelAlign{parright}{\parbox[t]{\labelwidth}{\raggedleft{#1}}}
\setlist[description]{style=multiline,topsep=4pt,align=parright}

\makeatletter
\let\reftagform@=\tagform@
\def\tagform@#1{\maketag@@@{(\ignorespaces\textcolor{black}{#1}\unskip\@@italiccorr)}}
\newcommand{\iref}[1]{\textup{\reftagform@{\tcr{\ref{#1}}}}}
\makeatother

\newenvironment{affiliations}{%
    \setcounter{enumi}{1}%
    \setlength{\parindent}{0in}%
    \slshape\sloppy%
    \begin{list}{\upshape$^{\arabic{enumi}}$}{%
        \usecounter{enumi}%
        \setlength{\leftmargin}{0in}%
        \setlength{\topsep}{0in}%
        \setlength{\labelsep}{0in}%
        \setlength{\labelwidth}{0in}%
        \setlength{\listparindent}{0in}%
        \setlength{\itemsep}{0ex}%
        \setlength{\parsep}{0in}%
        }
    }{\end{list}\par\vspace{12pt}}

\begin{document}

\title{Emergence of a Balanced Core through Dynamical Competition in Heterogeneous Neuronal Networks}
\author{Qing-long L. Gu\textsuperscript{1}, Songting Li\textsuperscript{2}, Wei P. Dai\textsuperscript{3}, Douglas Zhou\textsuperscript{1,\footnote{zdz@sjtu.edu.cn}}, David Cai\textsuperscript{1,2,4,\footnote{cai@cims.nyu.edu}}}

\date{}
\maketitle
\begin{affiliations}
 \item School of Mathematical Sciences, MOE-LSC, and Institute of Natural Sciences, Shanghai Jiao Tong University, Shanghai, China
 \item Courant Institute of Mathematical Sciences and Center for Neural Science, New York University, New York, NY, United States of America
 \item Department of Physics and Astronomy, and Institute of Natural Sciences, Shanghai Jiao Tong University, Shanghai, China
 \item NYUAD Institute, New York University Abu Dhabi, Abu Dhabi, United Arab Emirates\\
\end{affiliations}
\begin{abstract}
The balance between excitation and inhibition is crucial for neuronal computation. It is observed that the balanced state of neuronal networks exists in many experiments, yet its underlying mechanism remains to be fully clarified. Theoretical studies of the balanced state mainly focus on the analysis of the homogeneous Erd$\ddot{\text{o}}$s-R\'enyi network. However, neuronal networks have been found to be inhomogeneous in many cortical areas. In particular, the connectivity of neuronal networks can be of the type of scale-free, small-world, or even with specific motifs. In this work, we examine the questions of whether the balanced state is universal with respect to network topology and what characteristics the balanced state possesses in inhomogeneous networks such as scale-free and small-world networks. We discover that, for a sparsely but strongly connected inhomogeneous network, despite that the whole network receives external inputs, there is a small active subnetwork (active core) inherently embedded within it. The neurons in this active core have relatively high firing rates while the neurons in the rest of the network are quiescent. Surprisingly, the active core possesses a balanced state and this state is independent of the model of single-neuron dynamics. The dynamics of the active core can be well predicted using the Fokker-Planck equation with the mean-field assumption. Our results suggest that, in the presence of inhomogeneous network connectivity, the balanced state may be ubiquitous in the brain, and the network connectivity in the active core is essentially close to the Erd$\ddot{\text{o}}$s-R\'enyi structure. The existence of the small active core embedded in a large network may provide a potential dynamical scenario underlying sparse coding in neuronal networks.
\end{abstract}

\section*{Introduction}

Neuronal firing activity in the cortex can be highly irregular~\cite{britten1993responses,london2010sensitivity,compte2003temporally,shadlen1998variable}. Because the precise timing of spikes may contain substantial information about the external stimuli, irregular activity may serve as a rich encoding and processing space for neural computation~\cite{hertz1996learning,gutig2006tempotron,sussillo2009generating,monteforte2012dynamic}. To understand how the brain processes information, it is important to investigate how such irregularity emerges in the brain.

Some studies conclude that irregular firing may be regarded as noise, thus, conveying little information~\cite{shadlen1994noise,han2015optimum}. Meanwhile, other studies show that timing of spikes and the temporal activity patterns of  irregular neuronal firings \emph{in vivo} are able to convey specific information~\cite{richmond1990temporal,whalley2013neural,pillow2005prediction}. A germinating mechanism underlying irregular activity was proposed in the balanced network theory~\cite{van1996chaos,vreeswijk1998chaotic,troyer1997physiological,vogels2005neural,miura2007balanced}. In a balanced network, sparsely connected neurons possess strong architectural coupling but weak firing correlations between pairs of neurons. The excitatory and inhibitory inputs into each neuron, on average, are dynamically balanced with each other, suppressing the mean of the total input. Consequently, fluctuations of the input become dynamically dominant, giving rise to irregular firing events of each neuron. The defining characteristics of a balanced network includes a broad and heterogeneous distribution of the single-neuron firing rate and a linear response of the mean population firing rate to the external input. These properties have been extensively studied computationally during the development of the theory of balanced networks~\cite{vreeswijk1998chaotic,mehring2003activity,renart2010asynchronous}.  Consistent with theoretically predicted scenarios, certain experimental observations have been interpreted as consequences of balanced networks. For example, \emph{in vitro}, the sustained irregular activity of neurons in slices of the ferret prefrontal and occipital cortex was shown to be driven by the balance of proportional excitation and inhibition~\cite{shu2003turning}. \emph{In vivo}, the excitatory and inhibitory inputs to a neuron in ferret's prefrontal cortex were also found to be dynamically balanced ~\cite{haider2006neocortical}.

As shown in recent experimental data, the structure of developing hippocampal networks in rats and mice conforms to a scale-free (SF) topology, with the number of connections per neuron following a power-law distribution~\cite{bonifazi2009gabaergic}. Bidirectional and clustered three-neuron connection motifs were experimentally observed to occur with frequency significantly above chance in the visual system~\cite{song2005highly}, thus strongly deviating from statistically homogeneous networks. The network in the somatosensory cortex of neonatal animals was found to be a small-world (SW) network~\cite{perin2011synaptic}, that is, its connectivity has properties of high clustering and short average path lengths~\cite{newman2003structure}. In general, it is theoretically challenging to understand the dynamical consequences of these complex network architectures~\cite{boccaletti2006complex}.

Theoretical and computational works of the balanced state so far have mainly focused on the study of homogeneous random networks, \emph{i.e.}, those with the topology of the Erd$\ddot{\text{o}}$s-R\'enyi (ER) networks.  Because the network topology studied in the theory of balanced networks tends to be rather simplified, it is important to examine whether the current form of the balanced-state theory of homogeneous random networks~\cite{van1996chaos,vreeswijk1998chaotic,vogels2005neural,troyer1997physiological,vogels2005signal,miura2007balanced} can fully account for the balanced phenomena observed in experiments. As shown in a recent study that dynamics on a complex network can be controlled by the topology of the network~\cite{shkarayev2009architectural}, it is crucial to investigate the issue of how the topology of heterogeneous neuronal networks will affect the mechanism of the balanced state by inhomogeneity, for example, in the presence of groups of highly clustered neurons or hub neurons with a large number of presynaptic neurons. Finally, on account of the fact that the neuronal membrane potential was assumed to be a binary digital signal in the original theory~\cite{vreeswijk1998chaotic}, it behooves one to employ more realistic neuron models in the study of the balanced state in inhomogeneous networks.

To deepen our understanding of the balanced network and its potential application in neuronal networks arising from the brain, in this work, we address the questions of whether the existence of a balanced state sensitively depends on the network topology, what the characteristics of a  balanced state in an inhomogeneous network are, and what the dynamical implications of a balanced state for general complex networks are. Instead of the binary neuronal model, here, we choose the integrate-and-fire (I\&F) neuronal model. Our study leads to a novel characterization of a balanced network with complex topology: the balanced state always persists in a small active subnetwork, while the neurons in the rest of the network are quiescent. More specifically, the neuronal dynamics intrinsically drives the neuronal population into two subnetworks of distinct dynamics characterized by their firing rates: one subnetwork consisting of neurons that rarely fire --- which we refer to as the quiescent group, and the other subnetwork consisting of neurons that possess persistent irregular firing --- which we refer to as the active group. The subnetwork that contains all the active neurons and all the connections between the active neurons forms an active core embedded in the original inhomogeneous network, dominating the activity of the system. Surprisingly, the connectivity structure of the active core can be nearly characterized by an ER network and the active core exhibits the dynamics of the balanced state. The above phenomenon is confirmed for various inhomogeneous network topologies, including SW networks and SF networks with different degree-correlations as well as different exponents of degree distribution. As a consequence of the existence of this balanced active core, our results are different from previous studies about the dynamical consequences of heterogeneity in neuronal networks~\cite{landau2016impact, pyle2016highly}, and demonstrate that the balanced state might be ubiquitous for complex networks. The phenomenon of the active core appears consistent with a group of experimental observations related to sparse representation~\cite{o2010neural,hromadka2008sparse,poo2009odor,barth2012experimental}, that is, the information in each input is encoded by the firing of a relatively small set of neurons in the population. Such sparse activity has been observed in the barrel cortex of mice~\cite{o2010neural}, the auditory cortex of rats~\cite{hromadka2008sparse}, and the primary olfactory cortex of rats~\cite{poo2009odor}, elicited by a variety of stimuli. From this perspective, the active core may become an underlying dynamical substrate for sparse coding in neuronal networks.


\section*{Results}
\subsection*{Homogeneous balanced network theory}
To contrast with networks of other topologies below, we first recapitulate the balanced state in a homogeneous network, \emph{i.e.}, an ER network of binary neurons~\cite{vreeswijk1998chaotic}. An important feature of this balanced network is of sparse connection but strong coupling.  As discussed in Sec.~\emph{Materials and Methods} specifically, the average number of connections to each neuron from both presynaptic excitatory and presynaptic inhibitory populations is much smaller than the total number of neurons in the network, and the coupling strength is of the order $1/{\sqrt{K}}$. This scaling ensures persistent fluctuations of inputs in the large-$K$ limit. Below is a summary of the defining features of the balanced state in a homogeneous neuronal network:\\

\begin{figure}[!ht]
\centering
\includegraphics[width=1.0\textwidth]{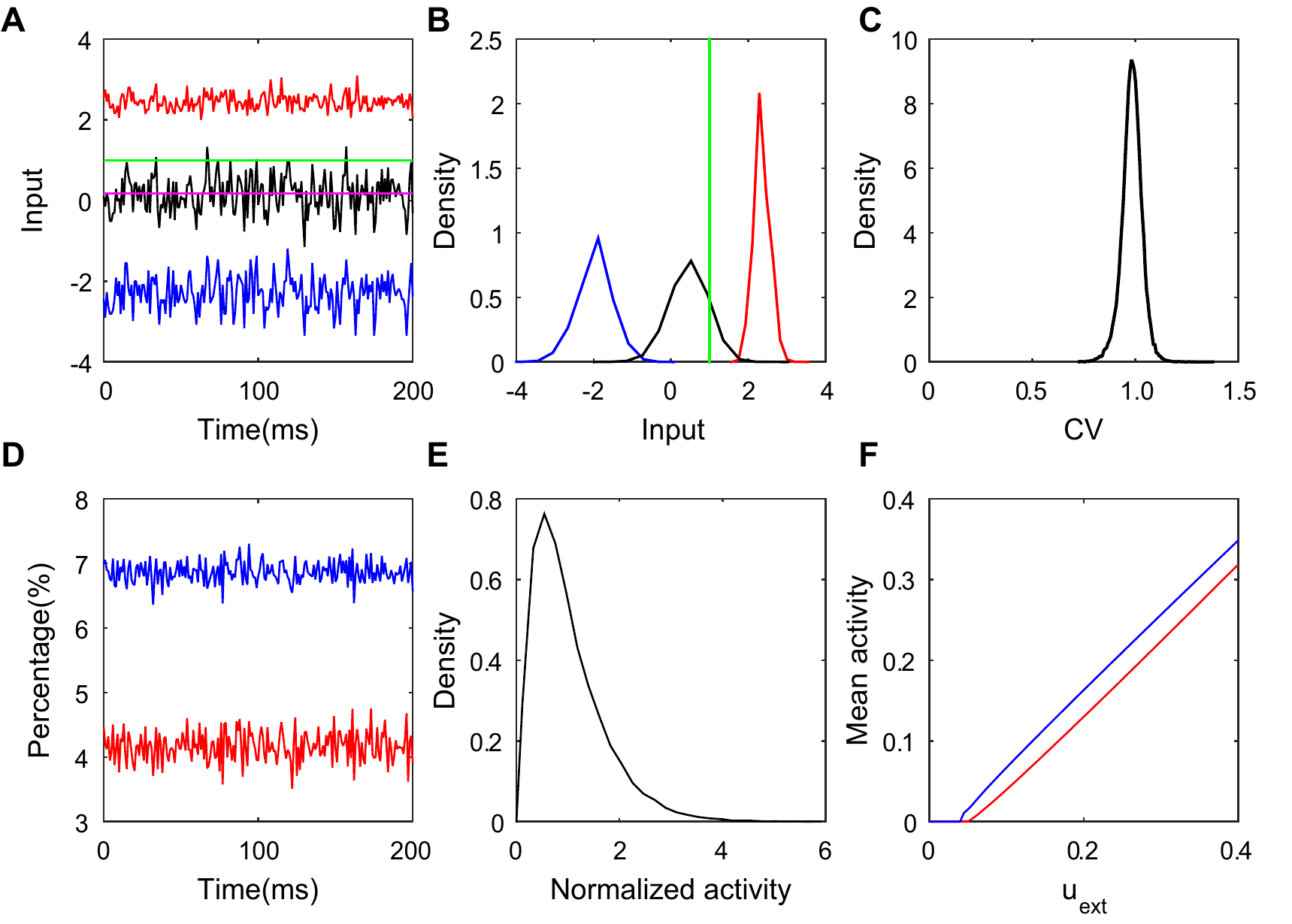}
\caption{{\bf Properties of a balanced network of binary neurons with homogeneous topology.} (A): The balanced excitatory and inhibitory inputs into a sample neuron (transient dynamics have been removed). The magnitudes of excitatory (red) and inhibitory (blue) inputs are greater than the firing threshold (green), whereas the total input (black) crosses the threshold stochastically with its mean (magenta, the value is $0.17$) remaining below the threshold; (B): The probability density functions of  the excitatory (red), inhibitory (blue) and total (black) inputs for the sample neuron in the panel (A). The green line stands for the threshold; (C): The distribution of the CV value for the ISIs of each neuron. The distribution is far from zero, indicating high firing irregularity of all neurons; (D): The population-averaged excitatory (red) and inhibitory (blue) activities. The population-averaged activity is the percentage of active neurons in the population at any given time (transient dynamics have been removed); (E): The distribution of the mean firing activity of each neuron; (F): The mean activity as a linear function of the external input parameter $u_{\text{ext}}$ for the excitatory population (red solid line) and the inhibitory population (blue solid line). Model details can be found in Sec. \emph{Materials and Methods}. Here, $N_E=N_I=2\times 10^4$ and $K=400$. In panels (A)-(E), $u_{\text{ext}}=0.1$.}
\label{fig1}
\end{figure}

\begin{enumerate}
\item{{\bf Balanced input:} As illustrated in Fig.~\ref{fig1}A, the magnitude of either the excitatory or the inhibitory input to a neuron is much higher than the neuronal firing threshold. However, because of the dynamically induced cancellation between the excitation and inhibition, the total input has its mean staying below the threshold all the time while its fluctuations stochastically driving the membrane potential across the threshold. Fig.~\ref{fig1}B illustrates the same phenomena from the viewpoint of distribution. The mean excitatory and inhibitory inputs of each neuron are large, while the mean total input of each neuron remains below the threshold;}\\

\item{{\bf Irregular activity:}  To quantify the irregular firing of a neuron in the balanced state, we use the coefficient of variation (CV), the ratio of standard deviation to mean, of the distribution of inter-spike-interval (ISI) for each neuron. Clearly, if $\text{CV} = 0$, the neuron fires regularly. As captured in Fig.~\ref{fig1}C, the distribution of the CV value is significantly far from zero for the balanced state;}\\

\item{{\bf Stationary population-averaged activity:} We next examine the population-averaged activity, $m_k(t)=\sum\limits_{i=1}^{N_k}{\sigma_k^i(t)}/ N_k$ for $k=E,I$, which is the percentage of active neurons in the population at any given time. For a balanced state, as shown in Fig.~\ref{fig1}D, the population-averaged activity almost stays constant over time as a consequence of the entire system reaching a stationary state;}\\

\item{{\bf Heterogeneity of firing rate:} As shown in Fig.~\ref{fig1}E, the activity of neurons is rather heterogeneous, which is characterized by a broad and rather skewed distribution of single-neuron firing rate in the network;}\\

\item{{\bf Linear response:} As exemplified in Fig.~\ref{fig1}F, a balanced state possesses the linear response of the mean activity of both the excitatory and inhibitory populations to the external input despite the nonlinear governing dynamics of each individual neuron. As noted above, in the large-$K$ limit, this defining property of linearity can be captured by Eq.~(\ref{eq:01line}) under the mean-field approximation.}\\
\end{enumerate}

All the above phenomena in the binary model can be explained analytically from the standpoint of the classical balanced network theory~\cite{vreeswijk1998chaotic}. Note that both the theory and simulations are based on the assumption that the network is homogeneous, \emph{i.e.} of the ER type, and that the neuron is of the binary type. These assumptions are highly simplified. Biological neuronal networks tend not to be homogeneous, \emph{e.g.}, the connections can be of SF~\cite{kaiser2007simulation,scannell1999connectional,sporns2004organization,sporns2007identification} or SW type~\cite{sporns2004small,sporns2006small,perin2011synaptic}. In general, it is expected that the topology could strongly influence the dynamics of neuronal networks~\cite{shkarayev2009architectural}. A natural and important extension of the theory is to examine the existence of a balanced state in heterogeneous networks. In the following, we first investigate the SF neuronal network, then discuss the case of the SW network. As an extension to the binary neuron model, we resort to the I\&F model in our simulation~\cite{carandini1996spike,rauch2003neocortical,zhou2009network,cai2005architectural,rangan2005modeling,zhou2013spatiotemporal}.

\subsection*{Uncorrelated SF network with I\&F neurons}

\begin{figure}[!ht]
\centering
\includegraphics[width=1.0\textwidth]{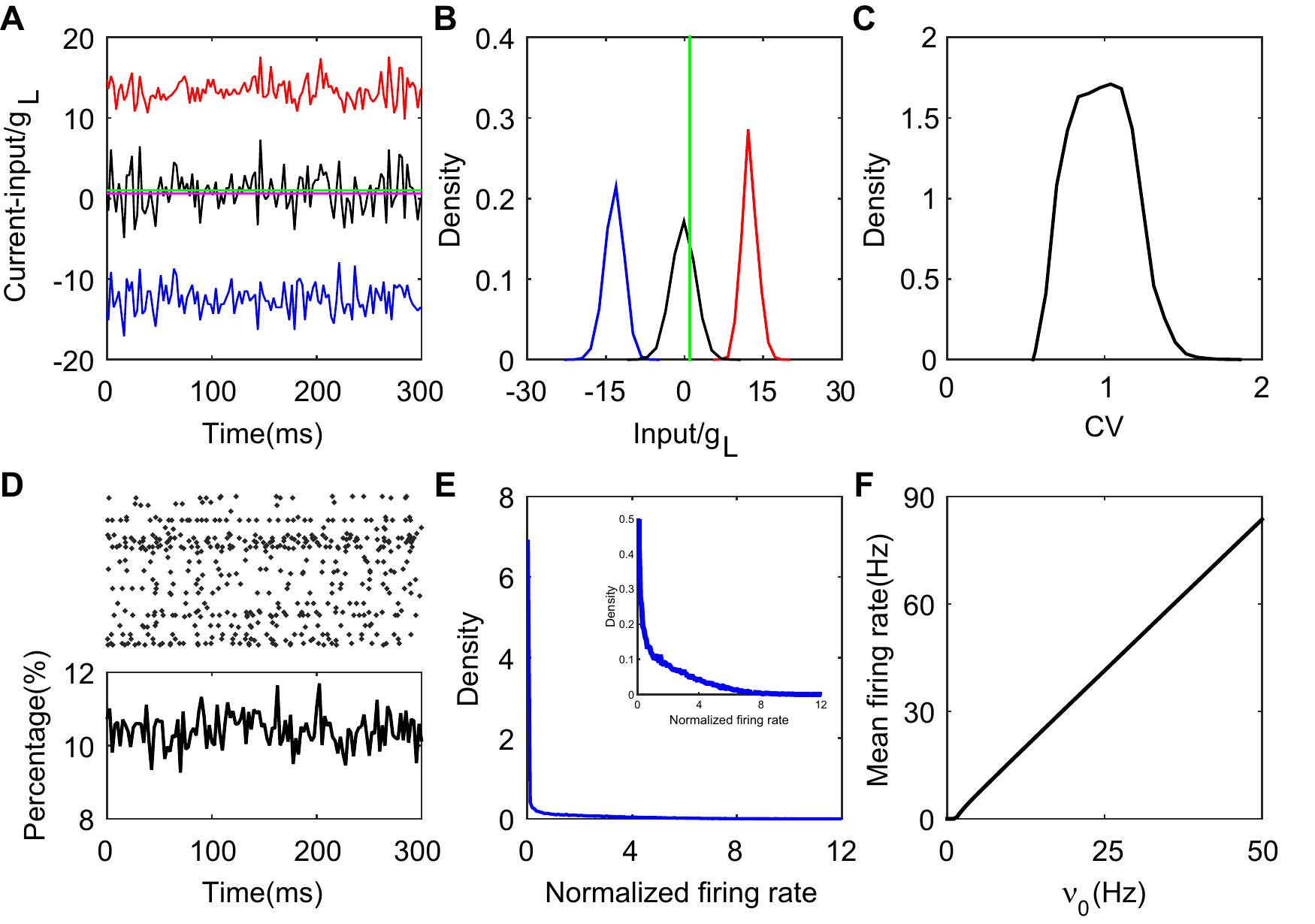}
\caption{{\bf Properties of an SF balanced network with pulse-current-based I\&F neurons.} (A): The balanced excitatory and inhibitory inputs into a sample neuron (transient dynamics have been removed). The magnitudes of the excitatory (red) and inhibitory (blue) inputs (scaled by the leakage conductance $g_L$) stay far away from the firing threshold (green), whereas the total input (black) (scaled by $g_L$) crosses the threshold stochastically with its mean (magenta, the value is $0.63$) remaining below the threshold; (B): The probability density functions of  the excitatory (red), inhibitory (blue) and total (black) inputs (scaled by $g_L$) for the sample neuron in the panel (A). The green line is the threshold; (C): The distribution of the CV value. Here, CV is calculated from the ISIs of each neuron; (D): The upper panel is the raster plot of a partial network (100 sample neurons selected at random from the network, with a time evolution of 300 ms), which exhibits asynchronous neuronal activity; the lower panel shows the percentage of the firing neurons over the network in each time window, where the time window is $2.5$ ms. The transient dynamics have been removed; (E): The distribution of neuronal firing rates (normalized by the mean firing rate averaged across the entire network). Inside we zoom in to the figure; (F): The mean firing rate of the excitatory and inhibitory populations as a linear function of the external input. Since we choose $J_{EE}=J_{IE}$ and $J_{II}=J_{EI}$, the gain curves for the excitatory and inhibitory populations overlap with each other. Here, $N_E=N_I=2\times 10^4$ and $K=400$. In panels (A)-(E), $\nu_0= 25$ Hz. }
\label{fig2}
\end{figure}

In this section, we address the question of whether there exists a balanced-network dynamics using the current-based I\&F neuronal model coupled with delta-pulse synaptic currents. This model is computationally simple but biologically more realistic than the binary model (the model details can be found in Sec. \emph{Materials and Methods}).

Here, we focus on the SF topology. We again invoke the coupling strength of order $1/{\sqrt{K}}$ to ensure that the network is fluctuation-driven when the mean connectivity $K$ is large. Under this scaling of coupling strength, our simulation results lead to the conclusion that an SF network can reliably evolve into a balanced state with its defining features. Illustrated in Fig.~\ref{fig2}A-B is the balance between the excitatory and inhibitory inputs to most of the neurons. For Fig.~\ref{fig2}A-B, we report the synaptic input at each moment by its time average within a small time window --- we select a time bin of 2.5 ms. Just as for neurons in the homogeneous balanced network, the CV value as shown in Fig.~\ref{fig2}C for the ISIs of each spiking neuron in the SF network is broadly distributed. This is consistent with the irregular activity of these neurons with heterogeneous connectivity. The population activity is asynchronous and stationary as the percentage of firing neurons fluctuates in time around a constant with a small amplitude (Fig.~\ref{fig2}D).  As exhibited in Fig.~\ref{fig2}E, strong heterogeneity is captured by the heavily skewed distribution of the single-neuron firing rate. Compared with the distribution of rates in the homogeneous system (Fig.~\ref{fig1}E), the firing rate distribution in the SF case manifests a sharp peak near the origin. We can conclude that there exists a group of neurons with extremely low firing rates or not firing at all (we will further discuss the significance of this phenomenon below). Finally, shown in Fig.~\ref{fig2}F is the linear response of both the excitatory and inhibitory populations to the external rate. To summarize, by the above defining characteristics of the balanced state, the stationary state in the SF I\&F neuronal network with delta-pulse synaptic currents can be readily identified as a balanced state. Next, we deploy the coarse-grained approach to further deepen our understanding of the dynamics in this SF system.

\subsection*{Quiescent and active groups}
For I\&F neurons with ER connections, under the homogeneity, any neuron in the population can be regarded as its representative in the mean-field sense. Recalling from Sec. \emph{Materials and Methods}, we have derived the Fokker-Planck (FP) equation (Eq.~(\ref{eq:FP-22})) for the population of ER connections. For a balanced state, we focus on the stationary solution of the FP equation for an ER network to obtain the linear relationship between the mean firing rate of the population and the rate of the external input~\cite{brunel2000dynamics}.

\begin{figure}[!ht]
\centering
\includegraphics[width=1.0\textwidth]{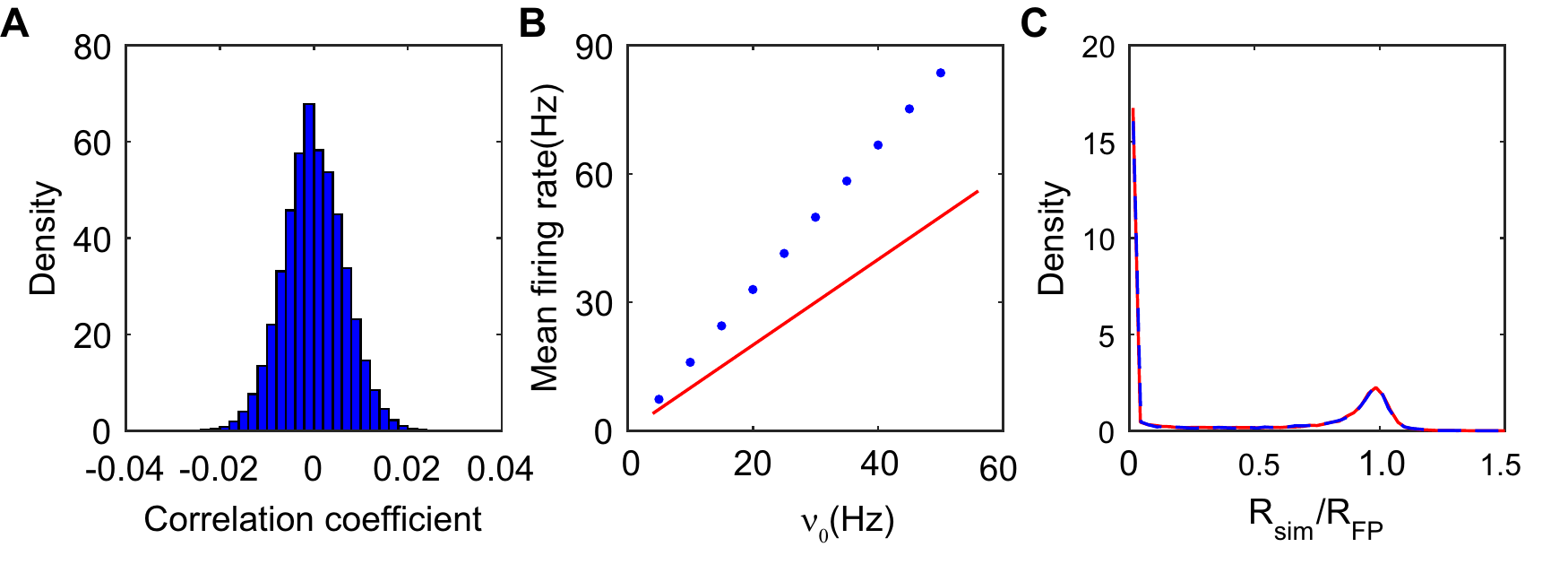}
\caption{{\bf FP description of an SF balanced network.} (A): The distribution of cross-correlation coefficient between spike trains of all pairs of neurons in the entire network. It is extremely narrowly centered around zero, therefore, the system is rather weakly correlated; (B): Gain curve. The red solid line is obtained from the FP equation with the mean-field assumption. Clearly, it strongly disagrees with that obtained from simulation (blue circles) which is the firing rate averaged over the entire network; (C): The distribution of ${R^i_{k,\text{sim}}}/{R^i_{k,\text{FP}}}$ across the excitatory (red solid line) and inhibitory (blue dashed line) populations. For the $i$th neuron in the $k$th population, $R^i_{k,\text{sim}}$ is the firing rate obtained from the simulation while $R^i_{k,\text{FP}}$ from the FP equation. The two curves overlap and have two peaks --- one is at zero and the other is near unity. All data is from the case discussed in Fig.~\ref{fig2}.}
\label{fig3}
\end{figure}

Next, we turn to an FP description of the neuronal dynamics in inhomogeneous networks. Recalling that the distribution of CV values of the ISI for a neuron's output spikes is not narrowly centered around unity as shown in Fig.~\ref{fig2}C, we thus cannot describe the output spike train of each neuron itself as a Poisson train in general. However, as shown in Fig.~\ref{fig3}A, the cross-correlation coefficient between the output spike trains from all pairs of neurons in the entire SF network is narrowly around zero with a magnitude of order $10^{-2}$. Therefore, the firing events for each pair of neurons are extremely weakly correlated. As a consequence,  the input into each neuron in the system can be regarded as three Poisson trains: the external, the excitatory, and the inhibitory synaptic inputs. Because of the sparse connections of the network and the absence of synchrony in the balanced state, these three Poisson trains are nearly independent of one another. Consequently, we can introduce the FP description just for the dynamics of a single neuron as Eq.~{\ref{eq:FP-1}.

As demonstrated above, the Poisson approximation is valid for any neuron with a given external input. However, because a neuron in an SF neuronal network cannot be regarded as an equivalent representative of all the neurons statistically, the mean-filed assumption ceases to be valid. Therefore, we cannot simply apply the same method used in the ER system as in Ref.\cite{brunel2000dynamics} (the FP equation with the mean-field assumption as Eq.~(\ref{eq:FP-22})) directly into the SF system, and our result shown in Fig.~\ref{fig3}B demonstrates that the solution of the FP equation under the mean-field assumption indeed fails to describe the linear population response property here. We then calculate the related rates $\{\nu^i_{kE}\}$ and $\{\nu^i_{kI}\}$ of the summed input spike trains to a neuron directly from the simulation for all the neurons in the network. Using these values of ${\nu^i_{kE}}$ and ${\nu^i_{kI}}$ in the FP equation (Eq.~(\ref{eq:FP-1})), we can obtain the firing rate $R_{k,\text{FP}}^i$ for the $i$th neuron in the $k$th population for the steady state. Meanwhile, we can also obtain the firing rate of the corresponding neuron from the simulation, which is denoted as $R_{k,\text{sim}}^i$. Furthermore, as shown in Fig.~\ref{fig3}C, there are clearly two peaks in the distribution of the ratio ${R_{k,\text{sim}}^i}/{R_{k,\text{FP}}^i}$ across the entire network. One peak is around unity and the other is around zero. The ratio near unity demonstrates that the activities of these neurons can be approximately captured by the description of the FP equation, while the zero ratio signifies the failure of the description of the FP equation. Here, the ratio for the $i$th neuron in the $k$th population equals zero only when $R_{k,\text{sim}}^i=0$, that is, the neuron fails to fire. As shown in Fig.~\ref{fig2}E, there indeed exists a large portion of the neurons that are nearly silent or do not fire at all. Interestingly, it was found experimentally that the firing activity in neocortical networks appears to be dominated by a small population of highly active neurons~\cite{yassin2010embedded}. Here, we find that the neuronal dynamics of SF networks separates the neuronal population into two subnetworks intrinsically: one consisting of neurons that fire no spikes, which will be referred to as the quiescent group; the other consisting of firing neurons, referred to as the active group (core). Subsequently, we study the difference between the quiescent and active groups, in particular, focusing on the properties of the active group.

\subsection*{Active core}
\begin{figure}[!ht]
\centering
\includegraphics[width=1.0\textwidth]{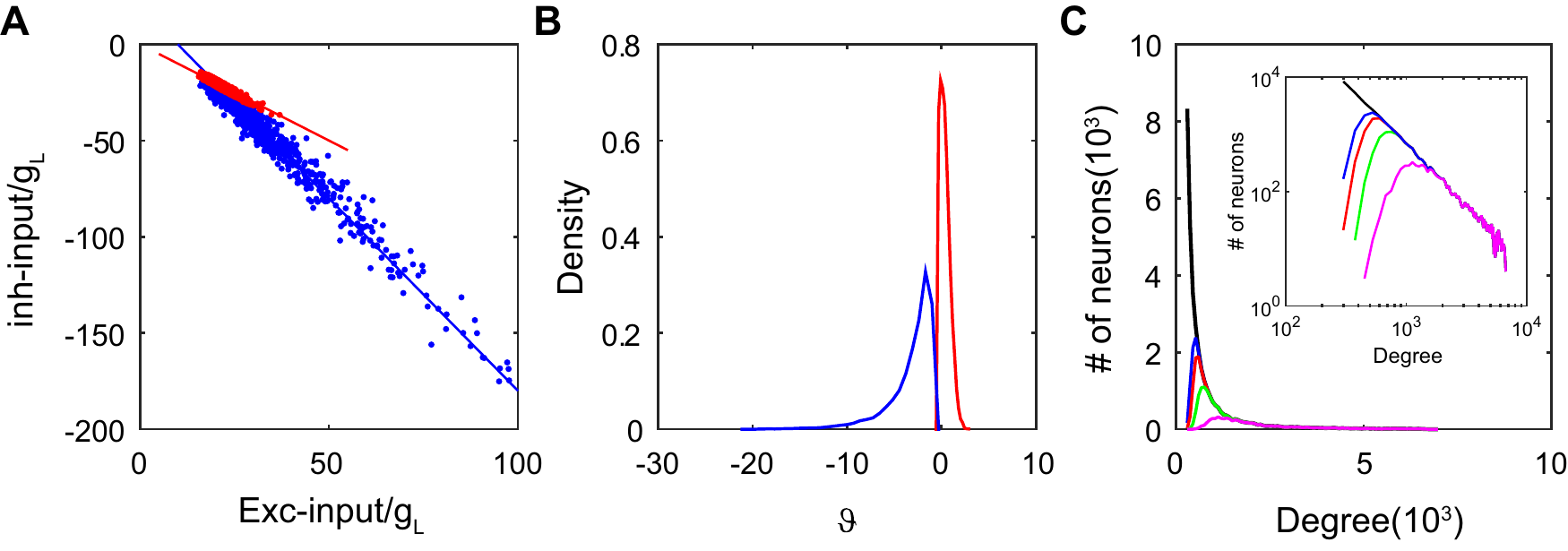}
\caption{{\bf Difference between the active and quiescent groups.} (A): The excitatory and inhibitory inputs (normalized by $g_L$) over the population. Each blue dot represents an inactive neuron, while each red dot represents an active neuron. Red and blue lines are linear fitting of the red and blue dots respectively. The slope is $-1$ for the red line while $-2$ for the blue line. Here, we select 1000 active and 1000 inactive neurons at random for the plot; (B): The distribution of $\vartheta$. $\vartheta$ is the time average of the total input into each neuron normalized by its standard deviation. The distribution of the active group (red) quantifies that the majority of neurons in the active group have fluctuation-dominated inputs (the mean nears zero), while in the quiescent group (blue) the magnitude of the total input is much greater than the standard deviation; (C): The degree distributions of the entire network (black solid line) and that of neurons in the quiescent group for different coupling strength ratio $\phi = J_{II}/{J_{EE}}$. The insert is the log-log plot for the same distributions. Here, $\phi=4$ for the blue solid line, $\phi=3$ for the red solid line, $\phi=2$ for the green solid line, and $\phi=1.2$ for the magenta solid line. The distributions agree with one another in the region of large degrees. Data in panels (A)-(B) is from the case shown in Fig.~\ref{fig2}.}
\label{fig4}
\end{figure}

We first examine whether there is any difference of inputs between the active and quiescent groups. From Fig.~\ref{fig4}A, it can be clearly observed that the quiescent group is strongly inhibited since the inhibitory input is twice larger in the quiescent group than that in the active group given the same excitatory input. By calculating the time-averaged total input to a neuron normalized by its standard deviation and denoting as $\vartheta$, it is clear from Fig.~\ref{fig4}B that the distribution of $\vartheta$ has a long negative tail for the quiescent group. Consequently, rarely can fluctuations drive their membrane potentials across the threshold. Note that the distribution of $\vartheta$ is concentrated around zero for the active group, thus indicating the neurons in the active group have fluctuation-dominated inputs. Next, we investigate the issue of how the coupling strength of the network affects the emergence of the active group. In particular, we focus on the competition between the excitatory and inhibitory coupling strength quantified by the ratio $\phi = J_{II}/J_{EE}$ with $J_{IE}=J_{EE}$ and $J_{EI}=J_{II}$. In our simulation, we fix the network topology while varying the value of $\phi$. Note that the degree distribution of the entire network is independent of $\phi$, since it is given by an SF network construction while the degree distribution of the active group depends on $\phi$ --- different coupling strengths give rise to different dynamics, which in turn generate different active cores dynamically. Fig.~\ref{fig4}C displays the degree distribution of the entire network and those of the quiescent groups with different values of $\phi$.  It is important to observe that these degree distributions agree with one another in the region of large degrees, that is, the quiescent group tends to be composed of the neurons with a large degree. This can be intuitively understood as follows. In the balanced network each neuron balances its external input with inputs from its presynaptic neurons. A neuron with a larger degree is likely to receive more excitatory and inhibitory synaptic inputs from other neurons. In addition, as exemplified in Fig.~\ref{fig4}A, the total input to a neuron, who receives strong excitatory and inhibitory synaptic inputs, tends to be sufficiently strongly inhibited to suppress its firing activity. Thus, for different values of $\phi$,  the neurons with high incoming degrees tend to belong to the quiescent group. Therefore, the main factor that separates the quiescent from the active groups in SF systems is the incoming degree distribution of a neuron. In general, the active group consists of neurons with low incoming degrees whereas the quiescent group consists of neurons with high incoming degrees. It is worthwhile pointing out that, with different network topologies and coupling strengths, the size of the active group can range from $10\%$ to $50\%$ of  the entire network.

We now focus on the subnetwork that contains only the active neurons and the connectivity structure of these neurons only. We will refer to this subnetwork as an \emph{active core}, which captures the spiking activity and the effective communication of the entire neuronal network. From Fig.~\ref{fig5}A, it is important to note that the degree distribution of the neurons in the active core is sharply peaked, resembling that of neurons in homogeneous networks.  Why does the degree distribution of the active core possess the characteristics of an ER network? For each neuron, we first examine the fraction of its active presynaptic neurons amongst all its presynaptic neurons, which will be denoted as $p$ below. The distribution of $p$ as shown in Fig.~\ref{fig5}B is sufficiently narrow to be approximated as a constant. Note that, for each neuron, $p$ can also be viewed as the probability of finding one of its presynaptic neurons to be active. The probability of finding a neuron with $w$ active presynaptic neurons can then be derived from the law of total probability, $P(w) = \sum_{k}{P(k)P(w|k)}$, where $P(k)$ is the probability of finding a neuron having $k$ presynaptic neurons, as the case here, whose distribution follows a power-law, $P(k) = ck^{-\gamma}$. By ignoring the correlation between the degree distribution of the active core and the formation of the active core, the conditional probability $P(w|k)$ can be approximated by a binomial  distribution $P(w|k)= C_k^w p^w (1-p)^{k-w}$. Further approximating the binomial distribution by a Gaussian, we can derive an approximation for $P(w)$:
\begin{equation}\label{eq:AGshape2}
P(w) \approx \sum_{k}{ck^{-\gamma}\frac{1}{\sqrt{2 \pi kp(1-p)}}\exp\Big{[}{-\frac{(w-pk)^2}{2kp(1-p)}}} \Big{]}.
\end{equation}
The probability $P(w)$ is a sum of a series of Gaussian terms with the coefficient of each term ordered by $k^{-\gamma}$. Therefore, a larger value of $k$ has a smaller contribution to the sum. In particular, for sufficiently large $\gamma$, the dominant term is a Gaussian.  As shown in Fig.~\ref{fig5}A, the prediction by Eq.~(\ref{eq:AGshape2}) is in very good agreement with the measured degree distribution of the active core.

\begin{figure}[!ht]
\centering
\includegraphics[width=0.7\textwidth]{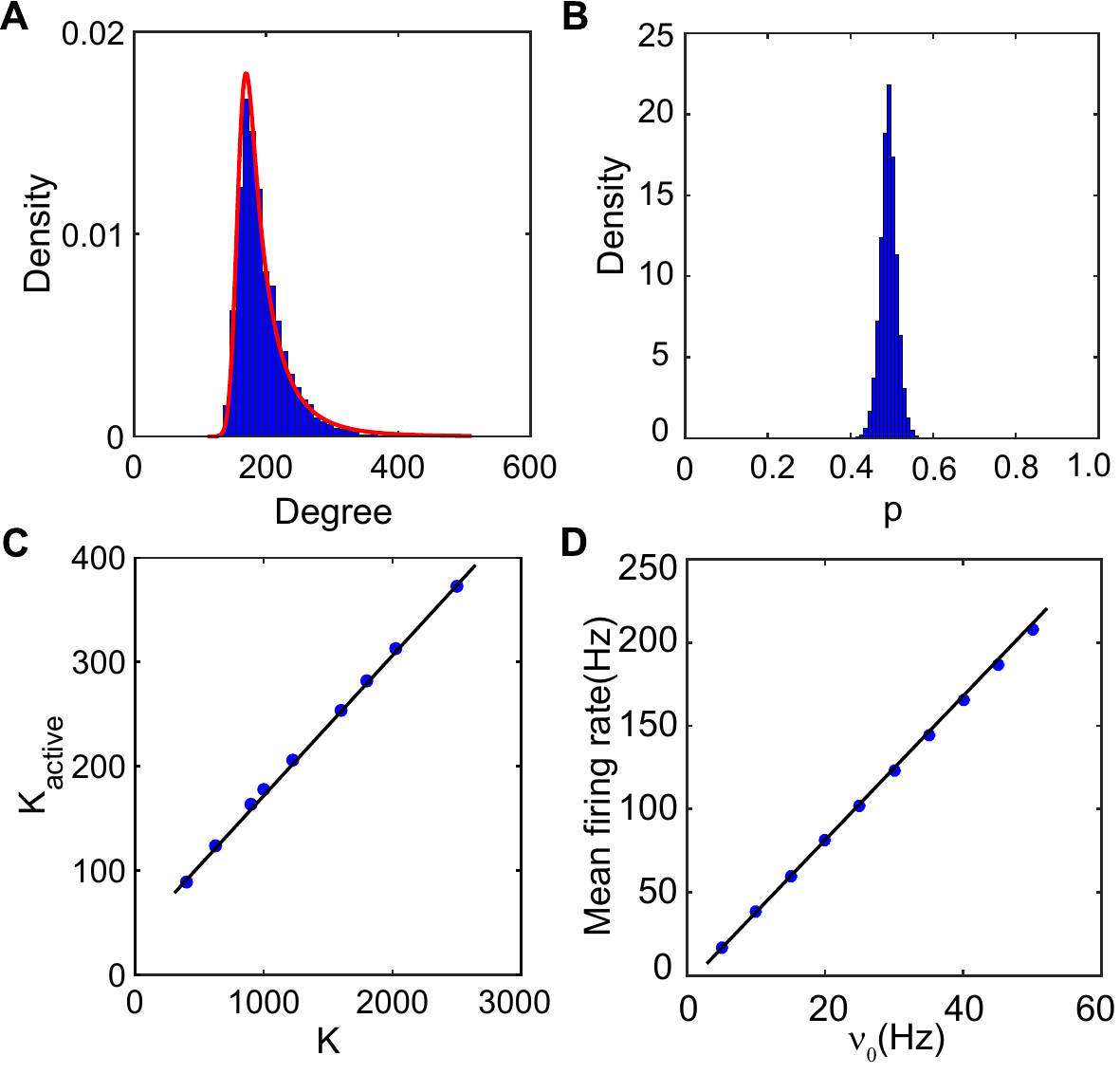}
\caption{{\bf Properties of the active core.} (A): The degree distribution in the active core. Numerical results (blue bars) can be well fitted by our prediction (Eq.~(\ref{eq:AGshape2}), red line). Note that the active core is the subnetwork consisting of all the active neurons and the connectivity structure of these active neurons; (B): The distribution of $p$ from the numerical simulation in the active core. For any single neuron, $p$ is the fraction of the number of its active presynaptic neurons over the number of its total presynaptic neurons. The distribution is narrowly centered around a constant; (C): Relationship between $K_{\text{active}}$ and $K$. The mean connectivity (degree) of the active core $K_{\text{active}}$ grows linearly with $K$. The black solid line is a linear fitting and the blue dots are results measured from the simulation; (D): The linear population response to the external drive in the active core. The linear population response from the simulation (blue dots) is in excellent agreement with the prediction by the FP analysis applied to the active core (black solid line). Data in (A), (B) and (D) is from the case shown in Fig.~\ref{fig2}.}
\label{fig5}
\end{figure}

Denoting the mean connectivity (in-degree) in the active core as $K_{\text{active}}$, we next examine the relationship between $K_{\text{active}}$ and $K$. Recall that $K$ is the average presynaptic connectivity of the original SF network. Numerically, we show that $K_{\text{active}}$ is proportional to $K$, as shown in Fig.~\ref{fig5}C, thus, when $K \rightarrow +\infty$, we also have $K_{\text{active}} \rightarrow +\infty$. Therefore, the dynamics of the active core possesses the same asymptotic behaviors as those of an ER network in the large-$K$ limit. Since the quiescent group does not influence the statistics of the irregular spiking events, the Poisson assumption of the summed presynaptic spike trains still holds in the active core. Using the mean-filed approximation for the active core, we further find that, as shown in Fig.~\ref{fig5}D, the linear response property of the active core is indeed well described by the FP analysis applied to the active core.

Note that the active core encompasses all spike events in the SF neuronal network, with connectivity similar to that of an ER network. The characteristics of the balanced state persists in the active core, that is the properties of \emph{balanced input}, \emph{irregular activity}, \emph{stationary population-averaged activity}, \emph{heterogeneity of firing rate}, and \emph{linear response} all hold. Clearly, our results demonstrate that the active core underlies the existence of the balanced state in the SF neuronal network.

\subsection*{Correlated SF networks with I\&F neurons}
Since the architectural degree-correlation may play an important role in the dynamics of a system~\cite{shkarayev2009architectural}, we generate SF networks with degree-correlation between neighbouring nodes using a reshuffling strategy~\cite{xulvi2005changing}. A balanced state can still arise in correlated SF neuronal networks. An example is shown in \ref{S1_fig} Fig, in which the five defining features of \emph{balanced input}, \emph{irregular activity}, \emph{stationary population-averaged activity}, \emph{heterogeneity of firing rate}, and \emph{linear response} again perseverate robustly. The degree distribution exponent $\gamma$ of the SF network used here is  $\gamma=2.6$, which is the same as that of the SF network for the uncorrelated case reported above (Fig.~\ref{fig2}). The degree correlation coefficient for the SF network in \ref{S1_fig} Fig is $\rho=0.03$. Similar to an SF network without degree correlation, the distribution of single neuron firing rates also possesses a high peak at zero. The SF network with degree correlation can also be decomposed into two subnetworks of distinct dynamics characterized by their firing rates. \ref{S2_Fig}E Fig demonstrates that the structure of the corresponding active core also displays that of homogeneous networks.

By generating SF networks with different correlation coefficients with $\gamma=2.6$, all of these SF systems exhibit the dynamics with a balanced active core. The degree distribution of the active core can be successfully described by Eq.~(\ref{eq:AGshape2}) for all values of $\rho$ ranging from $-0.3$ to $0.31$ as shown in \ref{S2_Fig} Fig. The properties of dynamics in these active cores are again similar to those of an ER balanced network.

In summary, our results confirm that the degree correlation between different nodes does not affect the properties of the balanced state in SF networks. For SF neuronal networks with degree correlations, the existence of the active core persists with the structure similar to that of an ER network and the active core possesses all the characteristics of the balanced state.

\section*{Discussion}

Many neuronal networks in the brain exhibit statistically nonhomogeneous connectivity structures. It has been observed that the connections of the neurons in layer 5 of the rat visual cortex display various highly clustered three-neuron connectivity patterns~\cite{song2005highly}. In addition, neuronal connectivity has been found to possess SF properties in rat hippocampal networks~\cite{bonifazi2009gabaergic}. The network connectivity between neurons in the somatosensory cortex of neonatal animals possesses the attributes of a small-world (SW) network~\cite{perin2011synaptic}. As can be expected, architectural properties play a crucial role in influencing the dynamics of neuronal networks~\cite{boccaletti2006complex}. Clearly, the question of how the topology influences the balanced state is of significant interest.

Our results have shown that a sparsely but strongly connected and uncorrelated SF network of I\&F neurons can still reach the balanced state. In addition to the pulse-coupled I\&F neurons, for the SF network of either binary neurons or smooth-current-based I\&F neurons, as shown in \ref{S3_Fig},~\ref{S4_Fig} and~\ref{S5_Fig} Figs, the balanced state also arises. These results imply that the balanced state is robust with respect to single neuron dynamics. Consistent with the recent works~\cite{pyle2016highly,landau2016impact}, by the distinct firing rate dynamics, the entire SF neuronal network can be driven into two subnetworks intrinsically: one is the quiescent group consisting of silent neurons and the other is the active group consisting of neurons with non-zero firing rates. Here, we emphasize that the subnetwork consisting of all the active neurons as well as the connectivity structure of these active neurons is further defined as the \emph{active core}. This active core possesses a degree-distribution characteristic of an ER network, which can be well described by our prediction (Eq.~(\ref{eq:AGshape2})), and displays similar dynamical properties of a balanced state of an ER network.

It has been shown that different architectural degree-correlations can induce different dynamical properties in scale-free networks~\cite{krapivsky2001organization}. These correlations can strongly influence the dynamics of the system~\cite{shkarayev2009architectural}. However, as far as the balanced state is concerned, there still exists the balanced state in SF neuronal networks with degree correlations, in which an ER-like active core controls its dynamics.

We have also examined the case of heterogeneous inputs in the simulation in addition to statistically identical Poisson inputs. \ref{S6_Fig}A-B Fig provides an example of the strength of the external input following a log-normal distribution~\cite{song2005highly} with a uniformly-distributed rate for different neurons. For this case, the dynamics still manifests a balanced active core whose in-degree distribution is again in excellent agreement with the prediction of Eq.~(\ref{eq:AGshape2}) shown in \ref{S6_Fig}C Fig.

As is shown that certain neuronal networks in the brain exhibit the SW characteristics~\cite{perin2011synaptic}, we have also conducted simulations with the SW connectivity. An active core of the balanced dynamics is again observed in the SW neuronal network with different rewiring probabilities (\ref{S7_Fig} Fig). The degree distribution in the active core is still close to that of an ER network. Our results suggest that the balanced state embedded in the active core may be ubiquitous for any heterogeneous networks.

Finally, we point out that the size of the active core found in our studies of heterogeneous neuronal networks ranges from $10\%$ to $50\%$ of the entire network, that is consistent with some experimental studies, which have shown that there exists a small embedded subnetwork of highly active neurons in different neocortex. For example, during a head-fixed object localization task, about half of all the neurons in a barrel column have been found to fire~\cite{o2010neural}. Experimental recordings in the primary auditory cortex of unanesthetized rats have shown that $50\%$ of the neural population failed to respond to any of the simple stimuli~\cite{hromadka2008sparse}. Furthermore, \emph{In vivo}, each odor can only evoke the activity of about $10\%$ neurons from anterior piriform cortex Layer 2/3~\cite{poo2009odor}. Thus the phenomenon of the active core in our results may provide insight into the potential mechanism for sparse coding.

\section*{Materials and Methods}

\subsection*{Degree distribution and degree correlation}
In the study of networks, the degree of a node in a network is the number of connections it has to other nodes. For a directed network, nodes have two different degrees, the in-degree, which is the number of incoming edges to a node, and the out-degree, which is the number of outgoing edges from a node. In this work, we mainly focus on the in-degree distribution, and just use degree instead of in-degree in the following for ease of discussion. The degree distribution $P(k)$ of a network is the probability of finding a node of degree $k$. The degree distribution of a directed ER network follows the Poisson distribution, $P(k)=\lambda^ke^{-\lambda}/k!$, which can be approximated by a Gaussian distribution for large $\lambda$ ($\lambda \gg 1$), $\lambda$ being the average degree of the network. The degree distribution of an SF network, by definition, follows a power-law distribution $P(k)\propto k^{-\gamma}$, $\gamma$ being the decay exponent~\cite{barabasi1999mean}.

Beyond the degree distribution, it is also important to characterize the degree-correlation between neighbouring nodes for large networks of complex structures~\cite{pastor2001dynamical, newman2003mixing}. In general, a network may display degree-correlations if the wiring probability between the high and low degree nodes statistically significantly differs from the independent random wirings between nodes. In our work, the degree-correlation is quantified by the Pearson correlation coefficient between the in-degrees for pairs of nodes linked by a directed edge.

\subsection*{The generation of scale-free neuronal networks}

To generate an SF network of size $N$, we first use the degree's power-law distribution to generate two numbers $k_i$ and $l_i$ to be the in-degree and out-degree, respectively, of the $i$th node, which are constrained by $\sum_i{k_i}=\sum_j{l_j}$. Then we choose at random a pair $(i,j)$ among these nodes. If the pre-existing in-degree of node $i$ is smaller than $k_i$, the pre-existing out-degree of node $j$ is smaller than $l_j$, and there is no directed edge from node $j$ to node $i$, we add a new directed edge from $j$ to $i$. Otherwise, we choose another pair at random to perform the previous procedure. We obtain an SF network by continuing this process until the in-degree of node $i$ equals $k_i$ while the out-degree of node $i$ equals $l_i$ for $i=1,2,...,N$. The degrees of the connected nodes in such an SF network are uncorrelated~\cite{aiello2000random,newman2001random}. We use a simple edge-node reshuffling strategy, which is a simplified version of the algorithm in Ref.~\cite{xulvi2005changing}, to generate an SF network with degree correlation.

Denoting $C_{kl}^{ij}$ as the element of the adjacency matrix with $C_{kl}^{ij}=1$ if there is a directed edge from the $j$th neuron in the $l$th population to the $i$th neuron in the $k$th population, where $k,l = E,I$. If each neuron is connected, on average, to $K$ presynaptic excitatory neurons and $K$ presynaptic inhibitory neurons, because each neuron is connected to a large number of presynaptic neurons in the cortex~\cite{braitenberg1998cortex}, the value of $K$ should be chosen sufficiently large to reflect this fact of connectivity. In addition, by electrophysiological recordings from cortical neurons, the probability of connection is shown to be often rather low, thus, yielding a sparse network~\cite{holmgren2003pyramidal}. Therefore, the value of $K$ should be chosen much smaller than the size of the population. As the cells in the primary visual cortex of adult cats were found experimentally firing much more irregularly \emph{in vivo} than the cells \emph{in vitro} when the same stimulus was used (passing the same current through the electrode), fluctuations of the synaptic inputs are particularly important for irregular spiking~\cite{holt1996comparison}. In this light, we choose the scaling of the coupling strength to be of order $1/\sqrt{K}$, imparting fluctuations of order one to persist in the large-$K$ limit in the total synaptic input to a neuron~\cite{van1996chaos,vreeswijk1998chaotic,vogels2005neural}. We adopt this scaling for all the neuron models used in this work.

\subsection*{The binary model}
In the binary model, the activity of the $i$th neuron in the $k$th population ($k=E,I$) is described by the binary variable $\sigma_{k}^{i}(t+\Delta t)=\Theta(u_{k}^{i}(t))$, where $\Theta(x)$ is the Heaviside function, and $u_{k}^{i}(t)$ equals the total synaptic input projecting into the $i$th neuron in the $k$th population above the threshold $\theta_{k}$ at time $t$,
\begin{equation}\label{eq:01input}
u_{k}^{i}(t)=I_{kE}^{i}(t) + I_{kI}^{i}(t)-\theta_{k},
\end{equation}
where $I_{kE}^{i}(t)=J_{kE}\sum\limits_{j=1}^{N_{E}}C_{kE}^{ij}\sigma_{E}^{j}(t) + u_k^0$ and $I_{kI}^{i}(t)=-J_{kI}\sum\limits_{j=1}^{N_{I}}C_{kI}^{ij}\sigma_{I}^{j}(t)$, $u_{k}^{0}$ is the constant external input to the $k$th population, and $J_{kl}$ descirbes the coupling strength from the $l$th population to the $k$th population ($k,l=E,I$), which is scaled as $1/\sqrt{K}$ as described above.

In our simulation, the values of the parameters chosen for the binary model are as follows : $J_{EE}=J_{IE}=1.0/\sqrt{K}$, $J_{II}=1.8/ \sqrt{K}$, $J_{EI}=2.0/{\sqrt{K}}$, $\theta_{E}=1.0$, $\theta_{I}=0.7$, $u_{E}^{0}=u_{\text{ext}}\sqrt{K}$, $u_{I}^{0}=0.8u_{\text{ext}}\sqrt{K}$, where $u_{\text{ext}}$ controls the magnitude of the external input.

For the balanced state in the ER network, one can obtain the mean population rate as~\cite{vreeswijk1998chaotic}
\begin{equation}\label{eq:01line}
m_{E}=\frac{1}{K}\frac{J_{II}u_{E}^{0}-J_{EI}u_{I}^{0}}{J_{EI}J_{IE}-J_{II}J_{EE}},\quad m_{I}=\frac{1}{K}\frac{J_{IE}u_{E}^{0}-J_{EE}u_{I}^{0}}{J_{EI}J_{IE}-J_{II}J_{EE}}.
\end{equation}
As mentioned above, both $u^0_E$ and $u^0_I$ are proportional to $u_{\text{ext}}$. Thus, Eq.~(\ref{eq:01line}) exhibits a linear relation between the mean activity rate $m_k$ and the external-input magnitude $u_{\text{ext}}$. This is one of the defining features of a balanced network.

\subsection*{The current-based I\&F model with delta-pulse coupling}
In our work, the sub-threshold membrane potential of an I\&F neuron in a population obeys the following dynamics~\cite{dayan2001theoretical,zhou2010spectrum,newhall2010dynamics}

\begin{equation}\label{eq:PIF}
\frac{dv_{k}^{i}}{dt}=-g_{L}(v_{k}^{i}-\epsilon_{R})+I_{k}^{i}(t),
\end{equation}
where $v_{k}^{i}$ is the membrane potential of the $i$th neuron in the $k$th population ($k=E,I$), $g_{L}$ is the leakage conductance, $\epsilon_{R}$ is the resting voltage, and $I_{k}^{i}(t)$ is the driving current. The voltage $v_{k}^{i}$ evolves according to Eq.~(\ref{eq:PIF}) while it remains below the firing threshold $\epsilon_{T}$. When $v_{k}^{i}$ reaches $\epsilon_{T}$, the $i$th neuron is said to fire a spike, and $v_{k}^{i}$ is set to the value of the reset voltage $\epsilon_{R}$. Upon resetting, $v_{k}^{i}$ is governed by Eq.~(\ref{eq:PIF}) again. At the same time, appropriate currents induced by the spike are injected into all other postsynaptic neurons. We use physiological values for the parameters $g_L = 50$ s$^{-1}$, $\epsilon_R=-70$ mV and $\epsilon_T=-55$ mV. Upon nondimensionalization, we have normalized $\epsilon_T=1.0$ and $\epsilon_R=0.0$.

The instantaneous current $I_{k}^{i}(t)$ injected into the $i$th neuron of the $k$th population has the following form $I_{k}^{i}(t)=I_{kE}^{i}(t)+I_{kI}^i(t)$, where $I_{kI}^i(t) = -J_{kI}\sum\limits_{j=1}^{N_{I}}C_{kI}^{ij}\sum\limits_{s}\delta(t-\tau_{js}^{I})$ is the inhibitory input, whereas $I_{kE}^{i}(t) = f_{k}\sum\limits_{s}\delta(t-\varsigma^k_{is})+J_{kE}\sum\limits_{j=1}^{N_{E}}C_{kE}^{ij}\sum\limits_{s}\delta(t-\tau_{js}^{E})$ is the excitatory input --- $\delta(\cdot)$ is the Dirac delta function, $J_{kl}$ is the coupling strength from the $l$th population to the $k$th population ($k, l= E, I$), and $f_{k}$ is the strength of the external Poisson input to the $k$th population. The first term in $I_{kE}^{i}(t)$ corresponds to the current from the external input. The external input of the $i$th neuron in the $k$th population is modeled by a Poisson process $\{\varsigma^k_{is}\}$ with rate $\nu_{k}$. At the time, $t=\varsigma^k_{is}$, of the $s$th input spike to the $i$th neuron in the $k$th population,  the neuron’s voltage jumps by the amount of $f_{k}$. The second term in $I_{kE}^{i}(t)$ and the term in $I_{kI}^{i}(t)$ correspond to the currents induced by the coupled neurons in the excitatory and inhibitory populations in the network, in which $\{\tau^E_{js}\}$ is the spike train from the $j$th neuron in the excitatory population, $\{\tau^I_{js}\}$ is the spike train from the $j$th neuron in the inhibitory population, and $s$ denotes the $s$th spike in the train.

In the simulation, the values of parameters in the model are set as follows: $J_{EE}=J_{IE}=1.0/{\sqrt{K}}$, $J_{II}= J_{EI}=2.0/{\sqrt{K}}$, $f_{E}=f_{I}=1.0/{\sqrt{K}}$, and $\nu_{E}=\nu_{I}=$ $\nu_{0}K$. We vary the value of $\nu_0$ to control the rate of the external input. To perform the numerical simulation of this I\&F model, we use an event-driven scheme~\cite{brette2007simulation}, with which the numerical results of dynamics can be obtained up to the machine accuracy.

\subsection*{Fokker-Planck equation}
Under a Poisson external input, the spiking events of a neuron in the network, in general, are not Poissonian, \emph{i.e.}, $\{\tau^E_{js}\}$ and $\{\tau^I_{js}\}$ in the current $I_k^i(t)$ are not a Poisson process for a fixed neuron $j$. However, the input to the $i$th neuron is a spike train summed over output spike trains from many other neurons in the network. If the firing event of each neuron is statistically independent of one another, then the spike train obtained by summing over a large number of output spike trains of neurons asymptotically tends to a Poisson process~\cite{cinlar1972superposition}. In a balanced network, the firing event of each neuron is extremely weakly correlated with, thus, nearly independent of other neurons~\cite{vreeswijk1998chaotic}. Therefore, for each neuron, the summed incoming spikes from its presynaptic neurons can be approximated by a Poisson train. Under the Poisson approximation, we can obtain the Fokker-Planck (FP) equation corresponding to Eq.~(\ref{eq:PIF}) for each neuron in the population~\cite{cai2006kinetic}. For the $i$th neuron in the $k$th population, we have
\begin{equation}\label{eq:FP-1}
\frac{\partial}{\partial t}\rho_k^i=\frac{\partial}{\partial v} \big{[}(g_{L}v-\mu_k^i)\rho_k^i \big{]}+\frac{{\sigma_k^i}^2}{2}\frac{\partial^{2}}{\partial v^{2}}\rho_k^i,
\end{equation}
where $\rho_k^i(v,t)$ is the probability density at time $t$ of finding the membrane potential at $v$ of the $i$th neuron in the $k$th population. Here $\mu_k^i$ is the mean total input,
\begin{equation}\label{eq:FP-1mean}
\mu_k^i=f_k\nu_k+J_{kE}\nu^i_{kE}-J_{kI}\nu_{kI}^i,
\end{equation}
and ${\sigma_k^i}^2$ is the strength of fluctuations of the total input,
\begin{equation}\label{eq:FP-1var}
{\sigma_k^i}^2=f_k^{2}\nu_k+J_{kE}^{2}\nu_{kE}^i+J_{kI}^{2}\nu_{kI}^i.
\end{equation}
Note that $\nu_{kE}^i$ and $\nu_{kI}^i$ are the rates of the summed respective excitatory and inhibitory inputs from other neurons in the network, $f_k$ and $\nu_k$ are the strength and rate of the external Poisson input to the $k$th population, respectively.

Equation~(\ref{eq:FP-1}) can be cast into the conservation form $\frac{\partial}{\partial t}\rho_k^i(v,t)+\frac{\partial}{\partial v}S_k^i(v,t)=0$, with $S_k^i(v,t)=-\frac{{\sigma_k^i}^2}{2}\frac{\partial}{\partial v}\rho_k^i-g_{L}\Big{(}v-\frac{\mu_k^i}{g_{L}}\Big{)}\rho_k^i$ being the probability density flux through $v$ at time $t$. For Eq.~(\ref{eq:FP-1}) we need to specify boundary conditions at $v=-\infty$, the reset potential $\epsilon_{R}$ , and the threshold $\epsilon_{T}$. The probability flux through $\epsilon_{T}$ gives the instantaneous firing rate at $t$, $m_k^i(t)=S_k^i(\epsilon_{T},t)$. For the I\&F neuron, its membrane potential cannot exceed the threshold, therefore, $\rho_k^i(v,t)=0$ for $v\geq\epsilon_{T}$. At the reset potential $v=\epsilon_{R}$ , there is a probability flux coming from the neuron that just crosses the threshold: what goes out at time $t$ at the threshold must come back at time $t$ at the reset potential, thus $S_k^i(\epsilon_{R}^+,t)-S_k^i(\epsilon_{R}^-,t)=m_k^i(t)$. The natural boundary condition at $v=-\infty$ is $\rho_k^i$ tending sufficiently rapidly toward zero to be integrable, ${\displaystyle \lim_{v\rightarrow-\infty}\rho_k^i(v,t)=0}$ and ${\displaystyle \lim_{v\rightarrow-\infty}v\rho_k^i(v,t)=0}$. By definition, $\rho_k^i(v,t)$ satisfies the normalization condition $\int_{-\infty}^{V_{T}}\rho_k^i(v,t)dv=1$.

As described previously, $K$ is chosen to be sufficiently large and the strength between neurons is scaled as $1/\sqrt{K}$, where $K$ is the average number of presynaptic connections in each population. Thus in Eq.~(\ref{eq:FP-1mean}), the scaling of the mean input $\mu_k^i$ has a leading order of $\sqrt{K}$. The physiological improbability of the value of $\mu_k^i$ tending to infinity in the large-$K$ limit, together with a balance between the excitatory and inhibitory inputs, forces the leading order of the right hand side in Eq.~(\ref{eq:FP-1mean}) to vanish, yielding the mean input $\mu_k^i$ to be of order one~\cite{vreeswijk1998chaotic,vogels2005signal}. The cancellation of the leading order in the FP description is a mathematical consequence of the balanced state.

For the balanced state in homogeneous neuronal networks, one can reach a probabilistic characterization of the network beyond the dynamics of a single neuron.  Because each neuron in the balanced state of a homogeneous network can be regarded as nearly statistically identical in a particular population, the input spike train of each neuron, which is summed from all presynaptic neurons, is Poisson with rate $K_lm_l(t)$, by noting that each neuron has $K_E$ presynaptic excitatory neurons and $K_I$ presynaptic inhibitory neurons on average. Here $m_l(t)$  is the population-averaged firing rate for a neuron in the $l$th population, $l=E,I$. Then, one can obtain
\begin{equation}\label{eq:FP-22}
\frac{\partial}{\partial t}\rho_k(v,t)+\frac{\partial}{\partial v}S_k(v,t)=0,
\end{equation}
where $\rho_k(v,t)$ is the probability of finding a neuron in the $k$th population whose membrane potential is $v$ at time $t$ ~\cite{brunel2000dynamics}, and the probability density flux $S_k(v,t) = -\frac{{\sigma_k}^2}{2} \frac{\partial}{\partial v}\rho_k$ $ - g_{L}(v-\frac{\mu_k}{g_{L}})\rho_k$, where the input is characterized by $\mu_k = f_k\nu_k+J_{kE}K_Em_E-J_{kI}K_Im_I$ and ${\sigma_k}^2 = f_k^2\nu_k+J^2_{kE}K_Em_E+J_{kI}^2K_Im_I$. By the same argument for Eq.~(\ref{eq:FP-1}), the boundary conditions for Eq.~(\ref{eq:FP-22}) can be similarly obtained.

\section*{Acknowledgement}
This work is supported by NYU Abu Dhabi Institute G1301 (Q.L.G., S.L.,W.P.D., D.Z., D.C.); NSFC-11671259, NSFC-11722107, NSFC-91630208, and Shanghai Rising-Star Program-15QA1402600 (D.Z.); NSFC-31571071 and NSF-DMS-1009575 (D.C.); Shanghai 14JC1403800, 15JC1400104 and SJTU-UM Collaborative Research Program (D.Z., D.C.).

\setcounter{figure}{0}
\renewcommand{\thefigure}{S\arabic{figure}}
\section*{Supporting Information}
\subsection*{S1 Fig}
\begin{figure}[!ht]
\centering
\includegraphics[width=1.0\textwidth]{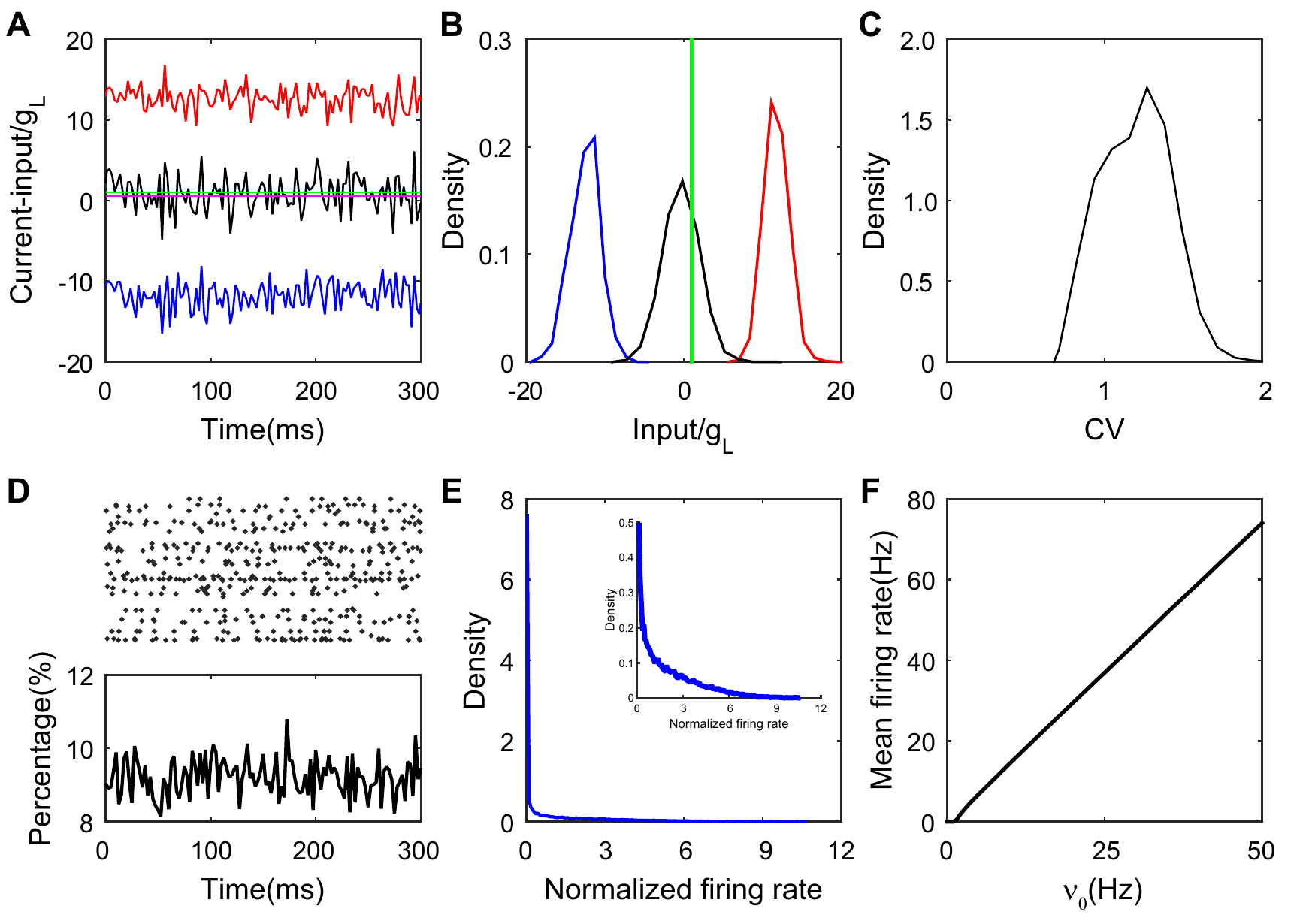}
\caption{{\bf Properties of a balanced SF neuronal network with degree-correlation.} (A): The balanced excitatory and inhibitory inputs of a sample neuron (transient dynamics have been removed). The magnitudes of the excitatory (red) and inhibitory (blue) inputs normalized by $g_L$ are much greater than the firing threshold (green), and the total input (black) normalized by $g_L$ has its mean (magenta, the value is $0.55$) below the threshold and intermittently crosses it; (B): The probability density functions of  the excitatory (red), inhibitory (blue) and total (black) inputs normalized by $g_L$ for the sample neuron in the panel (A). The green line is the threshold; (C): The distribution of CV value of a neuron's ISI over the network. It deviates significantly from zero as the firing activity in the neuronal network is rather irregular; (D): The upper panel is the raster plot of a partial network (100 sample neurons over 300 ms), showing that there is no synchrony; the lower panel shows the percentage of firing neurons in each time window remains nearly constant with small fluctuation over time as the system becomes stationary as time evolves. Here, the transient dynamics have been removed; (E): The distribution of single neuron mean firing rates normalized by the firing rate averaged across the network. Inside is the zoom-in of the figure; (F): The mean firing rate of the excitatory and inhibitory populations as a linear function of the external input. Since we choose $J_{EE}=J_{IE}$ and $J_{II}=J_{EI}$, the gain curves for the excitatory and inhibitory populations overlap with each other. Parameters here are the same as those in Fig.~\ref{fig2}. In panels (A)-(E), $\nu_0=25$ Hz. The degree correlation of the SF network is $\rho=0.03$.}
\label{S1_fig}
\end{figure}
{\bf Properties of a balanced SF neuronal network with degree-correlation.} The degree correlation coefficient of SF network is $\rho=0.03$. Our results in \ref{S1_fig} Fig show that there exists a balanced state in the correlated SF neuronal network.

\subsection*{S2 Fig}
\begin{figure}[!ht]
\centering
\includegraphics[width=1.0\textwidth]{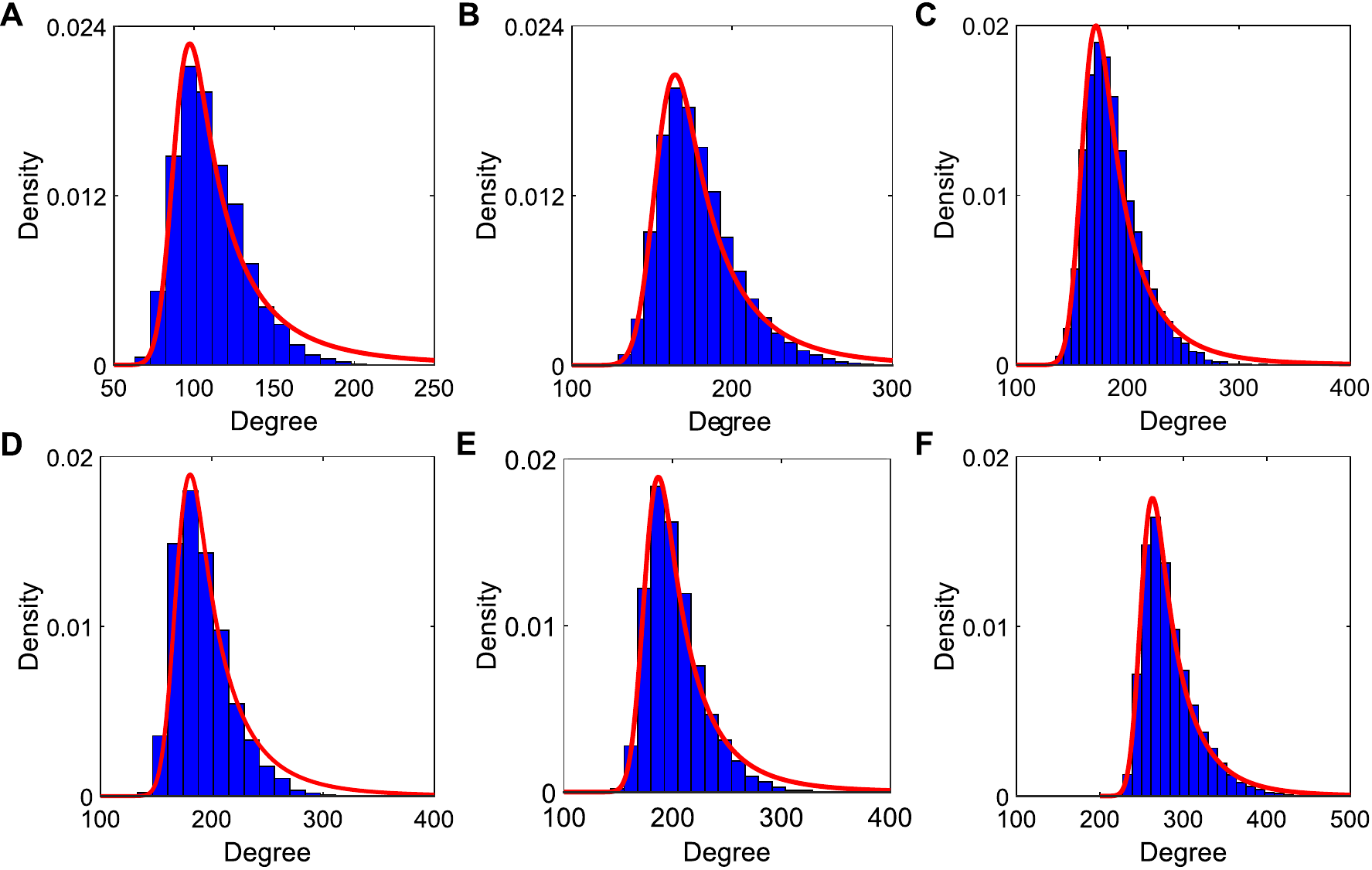}
\caption{{\bf Degree distribution of the active core in SF neuronal networks with different degree correlations.} The numerically measured distributions (blue bars) are well captured by the prediction of Eq.~(\ref{eq:AGshape2}) (red solid lines), (A): $\rho=-0.3$, (B): $\rho=-0.03$, (C): $\rho=-0.0042$, (D): $\rho=0.0036$; (E): $\rho=0.03$; (F): $\rho=0.31$.}
\label{S2_Fig}
\end{figure}
{\bf The active core in SF neuronal networks with different degree correlations.} The measured distribution of the active core in the SF neuronal network with different degree correlations, ranging from $\rho=-0.3$ to $\rho=+0.31$, is similar to that of an ER network. Note that the active core is the subnetwork consisting of all the active neurons and the connectivity structure of these active neurons.

\subsection*{S3 Fig}
\begin{figure}[!ht]
\centering
\includegraphics[width=1.0\textwidth]{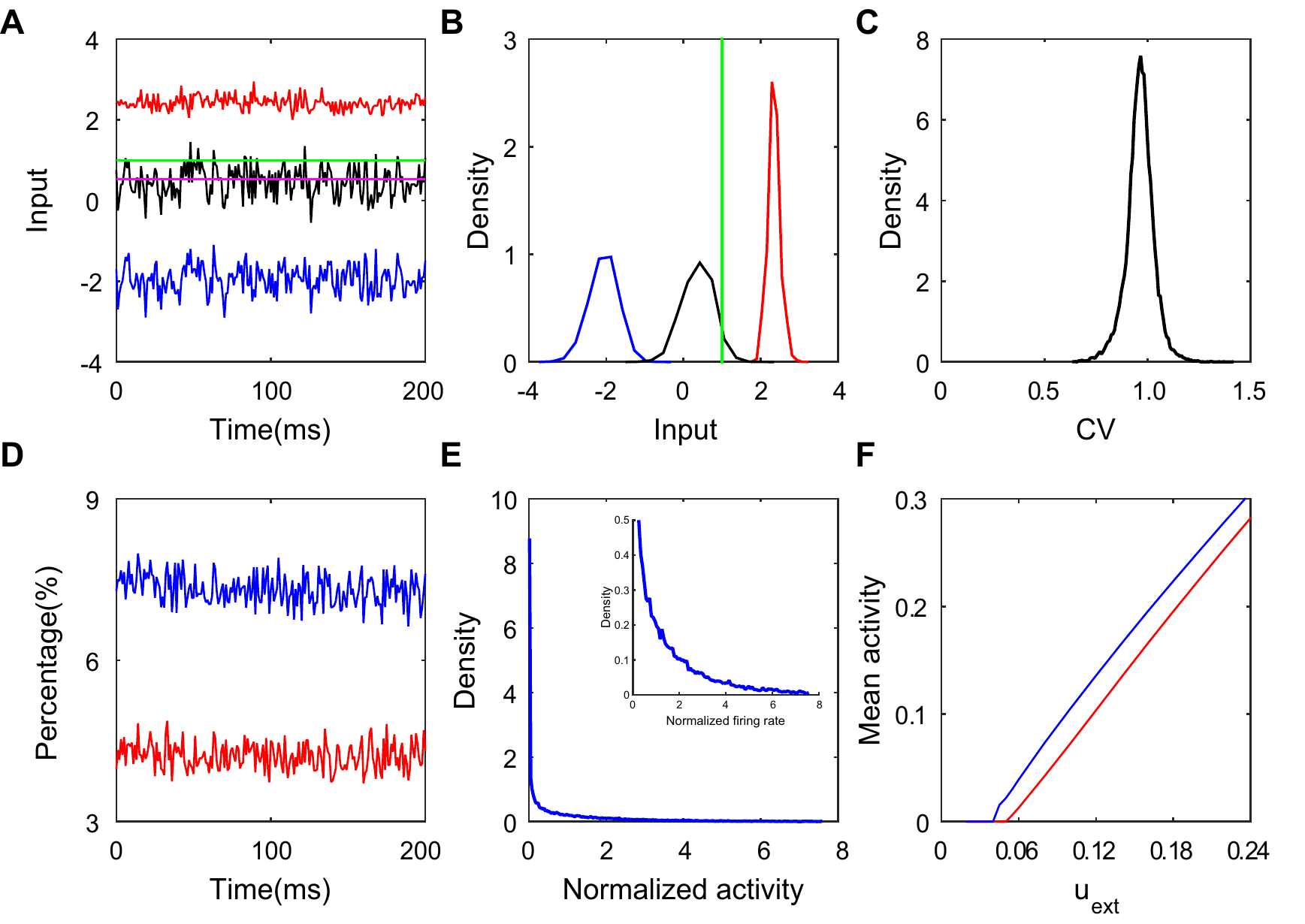}
\caption{{\bf Properties of an SF balanced network with binary neurons.} (A): The balanced excitatory and inhibitory inputs to a sample neuron (transient dynamics have been removed). The magnitudes of excitatory (red) and inhibitory (blue) inputs are larger than the firing threshold (green), while the mean (magenta, the value is $0.52$) of the total input (black) is always smaller than the threshold; (B): The probability density functions of the excitatory (red), inhibitory (blue) and total (black) inputs for the sample neuron in the panel (A). The green line is the threshold; (C): The distribution of the CV value. The CV is calculated from the ISIs of each neuron. It is far from zero, signifying the irregularity of neuronal activity; (D): The percentage of active neurons in the population at any given time. It signifies a stationary state by staying nearly constant with small fluctuations. The transient dynamics have been removed; (E): The distribution of the mean firing activity of each neuron normalized by the mean activity averaged across the network. Inside is the zoom-in of the figure. It exhibits heterogeneity of the firing rates; (F): The mean activity of the excitatory population (red solid line) and the inhibitory population (blue solid line) as a linear function of the external input parameter $u_{ext}$. Parameter values in this simulation are the same as those in Fig.~\ref{fig1}. In panels (A)-(E), $u_{\text{ext}}=0.08$. The SF network here is the same as that in the case shown in Fig.~\ref{fig2}.}
\label{S3_Fig}
\end{figure}
{\bf Binary model with SF connectivity.} We can find the balanced state in the SF network containing simple binary neurons.

\subsection*{S4 Fig}
\begin{figure}[!ht]
\centering
\includegraphics[width=1.0\textwidth]{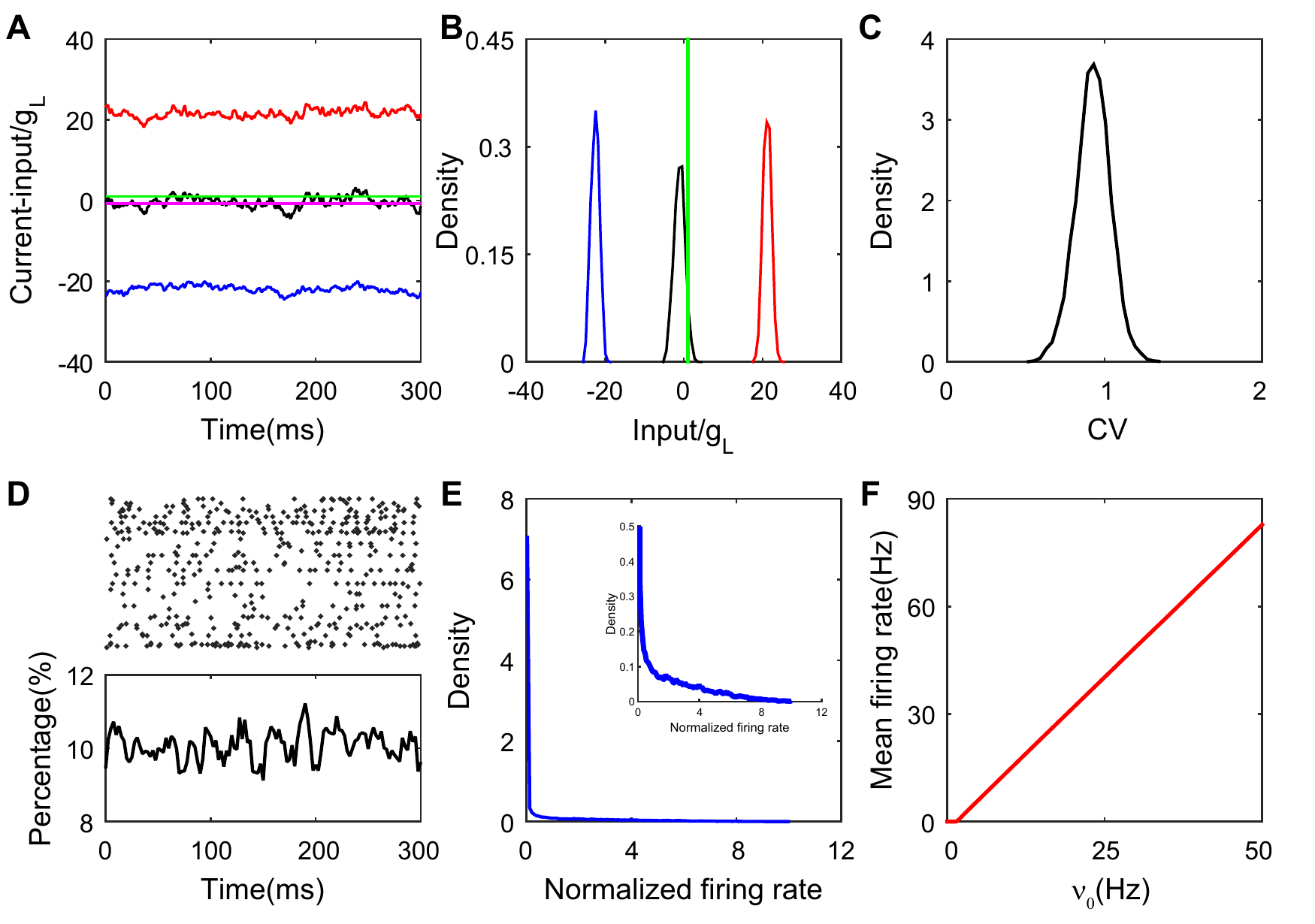}
\caption{{\bf Properties of an SF balanced network with smooth-current-based I\&F neurons.} (A): The balanced excitatory and inhibitory inputs of a sample neuron (transient dynamics have been removed). The magnitudes of excitatory (red) and inhibitory (blue) inputs normalized by $g_L$ are much larger than the firing threshold (green). Due to the cancellation between excitation and inhibition, the total input normalized by $g_L$ (black) has a small amplitude and occasionally crosses the threshold. The magenta line indicates the mean (the value is $-0.69$) of the total input; (B): The probability density functions of the excitatory (red), inhibitory (blue) and total (black) inputs (normalized by $g_L$) for the sample neuron in the panel (A). The green line is the threshold; (C): The distribution of the CV value of a neuron's ISI over the network. Since it is far from zero, all spiking neurons in the network fire irregularly; (D): The upper panel is the raster plot of a partial network (100 sample neurons over 300 ms), which exhibits asynchronous neuronal activity; the lower panel shows the percentage of the neurons that spikes over the population in each time window, where the time window is 2.5ms. The percentage of the spiking neurons in each time window almost keeps constant in time. The transient dynamics have been removed; (E): The distribution of neuronal firing rates normalized by mean firing rate averaged across the network. Inside is the zoom-in of the figure. The distribution is heavily skewed, exhibiting a strong variability of firing rate; (F): The mean firing rate of the excitatory and inhibitory populations as a linear function of the external input. Since we choose $J_{EE}=J_{IE}$ and $J_{II}=J_{EI}$, the gain curves for the excitatory and inhibitory populations overlap with each other. Parameters here are the same as those in Fig.~\ref{fig2}. In panels (A)-(E), $\nu_0=25$ Hz. The SF network here is the same as that in the case shown in Fig.~\ref{fig2}.}
\label{S4_Fig}
\end{figure}
{\bf Smooth-current-based I\&F neuronal network with SF connectivity.}  In this model we use $\alpha(t)= \Big{(}\exp{(-\frac{t}{\tau^k_{r}})}-\exp{(-\frac{t}{\tau^k_{d}})}\Big{)}/({\tau^k_{r}-\tau^k_{d}})$ (for $k=E, I$) as the smooth current input instead of delta-pulse current input and the sub-threshold membrane potential of a neuron still obeys Eq.~(\ref{eq:PIF}). Here,  $\tau_r^E=\tau_r^I = 1\,\text{ms}$, $\tau_d^E=5\,\text{ms}$, $\tau_d^I=10\,\text{ms}$ .

\subsection*{S5 Fig}
\begin{figure}[!ht]
\centering
\includegraphics[width=0.80\textwidth]{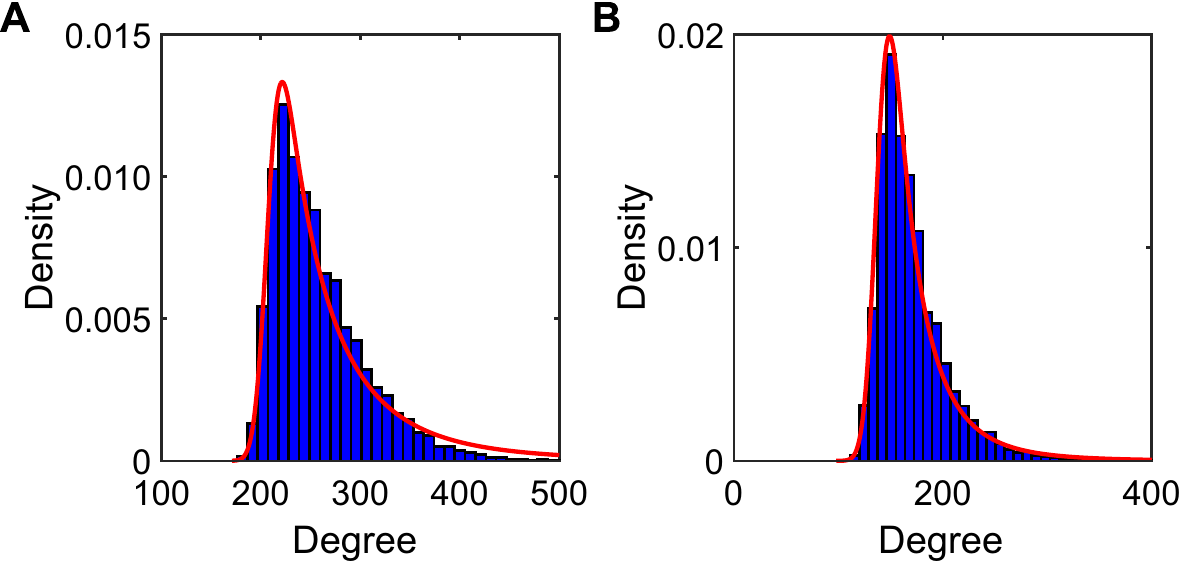}
\caption{{\bf Degree distribution of the active core in the SF network of different type of single neuron models.} The measured distribution (blue bars) of the active core for the SF network consisting of binary neurons in (A) and current-based I\&F neurons with smooth coupling in (B) is well captured by the prediction of Eq.~(\ref{eq:AGshape2}) (red solid lines). Here the active core is the subnetwork consisting of all the active neurons and the connectivity structure of these active neurons. Data in the panel (A) is from the case shown in \ref{S3_Fig} Fig; data in the panel (B) is from the case shown in \ref{S4_Fig} Fig.}
\label{S5_Fig}
\end{figure}
{\bf The degree distribution of the active core for the SF network consisting of binary neurons and smooth-current-based I\&F neurons.} The measured distribution of the active core from the SF network consisting either of binary neurons or of smooth-current-based I\&F neurons is similar to that of an ER network.

\subsection*{S6 Fig}
\begin{figure}[!ht]
\centering
\includegraphics[width=1.0\textwidth]{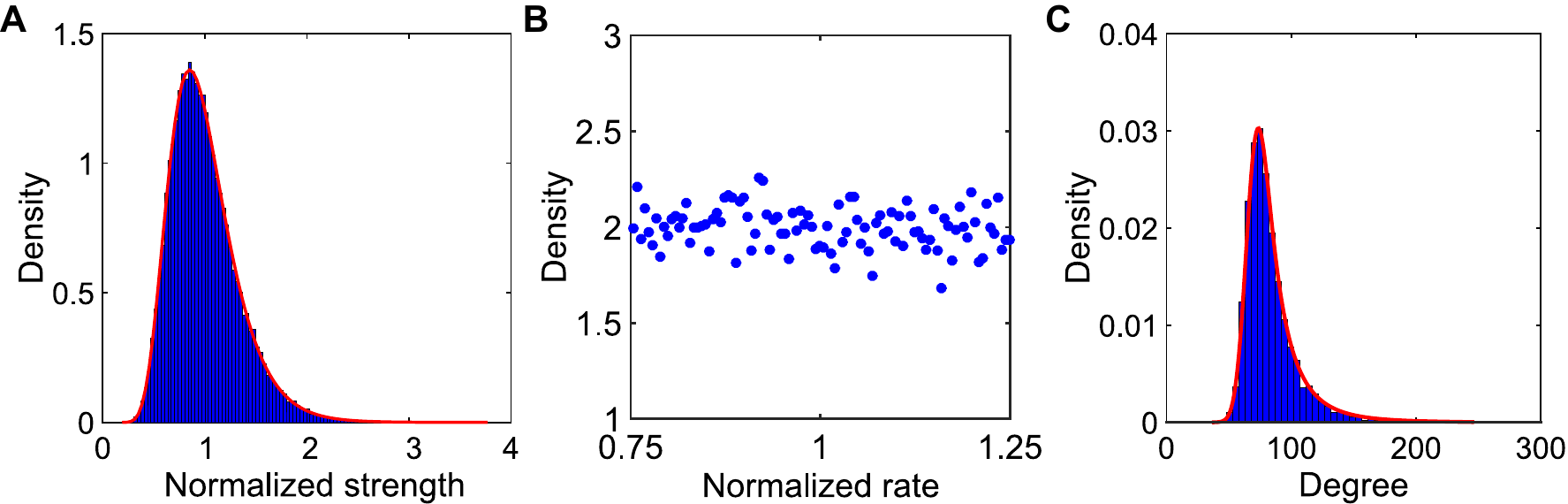}
\caption{{\bf Heterogeneous input} (A): The strength distribution of the external input normalized by the mean strength value averaged across the entire network. The simulation data (blue) is fitted by the log-normal distribution(red line); (B): The rate distribution of the external inputs normalized by the mean rate value averaged across the entire network. The rate (blue dot) in our simulations is uniformly distributed; (C): The degree distribution in the balanced active core of  the SF neuronal network with heterogeneous inputs. Measured distribution (blue bars) agrees well with our prediction of Eq.~(\ref{eq:AGshape2}) (red line). Here we use an SF network with degree correlation $\rho=0$.}
\label{S6_Fig}
\end{figure}
{\bf Heterogeneous input with SF connectivity.} We consider a case of using a heterogeneous input into the SF neuronal network. The strength of the external input follows a log-normal distribution. The rate of the external input follows a uniform distribution.

\subsection*{S7 Fig}
\begin{figure}[!ht]
\centering
\includegraphics[width=1.0\textwidth]{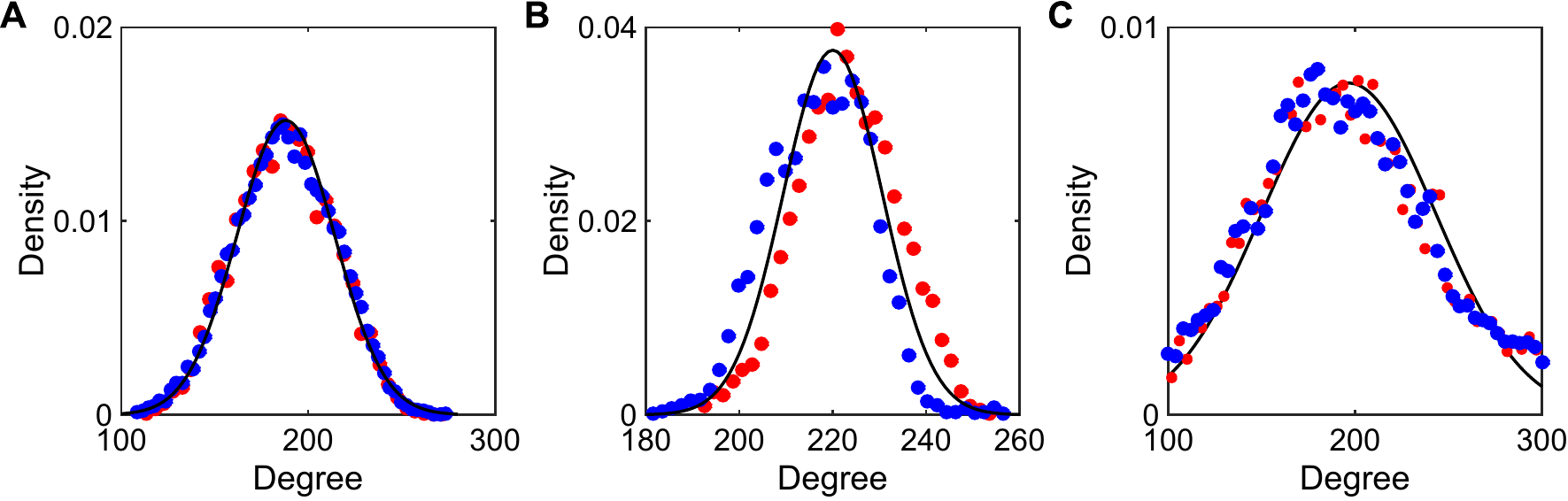}
\caption{{\bf Active cores in SW networks with different rewiring probabilities}  (A)-(C): The degree distribution in the active core. Simulation data (dots) compares with the Gaussian distribution (solid line). Red (blue) dots are for connections from presynaptic excitatory (inhibitory) neurons in the active core. The network is generated with the rewiring probability $q = 10^{-1}$ in (A), $q = 10^{-2}$ in (B) and $q = 10^{-3}$ in (C)}
\label{S7_Fig}
\end{figure}
{\bf Active cores in SW networks with different rewiring probabilities.} We generate the SW network with the following distance-dependent probability: the connected probability between $i$th and $j$th neuron is $p_{ij} = q p_0 + (1-q) \Theta(p_0-d_{ij})$, where $d_{ij} = \min(|i-j|,N-|i-j|)/(N/2)$, $N$ is the total number of neurons in the network, $p_0$ is the sparsity and $q$ is the rewiring probability~\cite{song2014simple}. There always exists a balanced active core in neuronal networks with different rewiring probabilities $q$.

\end{document}